# Automatic Detection of Omega Signals Captured by the Poynting Flux Analyzer (PFX) on Board the Akebono Satellite

I Made Agus Dwi Suarjaya, Yoshiya Kasahara, and Yoshitaka Goto
Graduate School of Natural Sci. and Tech.
Kanazawa University,
Kakuma-machi, Kanazawa 920-1192, Japan

*Abstract*—The Akebono satellite was launched in 1989 to observe the Earth's magnetosphere and plasmasphere. Omega was a navigation system with 8 ground stations transmitter and had transmission pattern that repeats every 10 s. From 1989 to 1997, the PFX on board the Akebono satellite received signals at 10.2 kHz from these stations. Huge amounts of PFX data became valuable for studying the propagation characteristics of VLF waves in the ionosphere and plasmasphere. In this study, we introduce a method for automatic detection of Omega signals from the PFX data in a systematic way, it involves identifying a transmission station, calculating the delay time, and estimating the signal intensity. We show the reliability of the automatic detection system where we able to detect the omega signal and confirmed its propagation to the opposite hemisphere along the Earth's magnetic field lines. For more than three years (39 months), we detected 43,734 and 111,049 signals in the magnetic and electric field, respectively, and demonstrated that the proposed method is powerful enough for the statistical analyses.

*Keywords—Auto-detection; Satellite; Signal processing; Wave Propagation; Plasmasphere*

## I. INTRODUCTION

The Akebono (EXOS-D) satellite was launched at 23:30 UT on February 21, 1989 to observe the Earth's magnetosphere and plasmasphere. This satellite has onboard VLF instruments and Poynting flux Analyzer (PFX) is one of the subsystems. The PFX is a waveform receiver that measures two components of electric fields and three components of magnetic fields with a band-width of 50 Hz in a frequency range from 100 Hz to 12.75 kHz [1]. Omega was a navigation system with 8 ground stations transmitter and had transmission pattern repeating every 10 s. Each station transmitted a different pattern of frequency but had a common frequency at 10.2 kHz. The Omega system was terminated in 1997 in favor of the GPS system.

From 1989 to 1997, the PFX on board the Akebono satellite received signals at 10.2 kHz from the eight stations and huge amounts of PFX data became valuable to study the propagation characteristics of VLF waves in the ionosphere and plasmasphere. Once the Omega signals are radiated in the plasmasphere permeating through the ionosphere, they propagate as a whistler mode wave in the plasma. The omega signal data captured by the PFX on board the Akebono has been used to estimate global plasmaspheric electron density. In particular, a tomographic electron density profile could be determined by calculating the Omega signal propagation path using the ray tracing method. This method could estimate the propagation path within one hour of single satellite observations [2]. The algorithm was further improved with a flexible method and novel stochastic algorithm. This enabled estimates separate from the effects of the ionosphere and plasmasphere [3]. Goto et al. (2003) demonstrated that electron density profile could be drastically changed day by day depending on magnetic activities during a magnetic storm.

Because the transmission pattern of frequency, time, and the location of each station is known, we can easily distinguish the signal source. We can then determine many propagation properties such as attenuation ratio, propagation direction, propagation time (delay time) from the transmission station, and the observation point along the satellite trajectories. Such parameters depend strongly on the plasma parameters along the propagation path. Therefore it is worth to analyze such propagation properties statistically using the long term observation data.

Processing manually all of the data from 1989 to 1997, however, will take a lot of time. Furthermore, more processes and analyses will be needed to see different results such as analysis based on magnetic local time, seasonal propagation analysis, and yearly propagation analysis. Our automatic detection method makes all of the analysis processes simpler and is able to produce most of the required result faster and efficient. This study discusses the automatic detection methods for faster analysis of huge amounts of PFX data to study the propagation characteristics of VLF waves which, in this case, is the Omega signal. An outline of the paper follows. In the second section, we present the technique we are using for automatic detection. It involves identifying a transmission station, calculating the delay time, and estimating the signal intensity. In the third section, we discuss the result of this analysis using an event study and statistical study for the Norway station. In the final section, we present conclusions and summarize our research.

## II. DATA ANALYSIS TECHNIQUE

### A. Omega Signal

The Omega signal is very low frequency (VLF) signal between 10 and 14 kHz transmitted by the Omega navigation system that was operational in 1971. Before it shut down in 1997, its purpose was to provide a navigational aid for domestic aviation and oceanic shipping. Omega receiver determines location based on the phase of the signal from two or more of the Omega stations [5]. This Omega signal was transmitted from eight ground stations with each station transmitting a unique pattern, based on which our analyzer software could determine the source of the signal. The location of these Omega stations are Norway (NW), Liberia (LB), Hawaii (HW), North Dakota (ND), La Reunion Island (LR), Argentina (AZ), Australia (AS), and Japan (JP). There are four common frequencies of the Omega signal (10.2, 13.6, 11.33, and 11.05 kHz) with one unique frequency for each station. The transmission pattern at 10.2 kHz from each station is shown in Fig. 1. Each station transmit 10.2 kHz signal at different timing. There is an interval of 0.2 s separating each of the eight transmissions, with variations in the duration for each station.

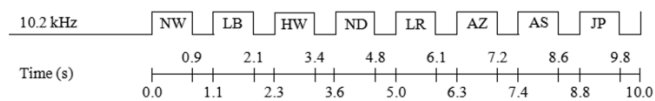

Fig. 1. Transmission Pattern of the 10.2 kHz Omega Signal. There are four common frequencies of the Omega signal (10.2, 13.6, 11.33, and 11.05 kHz) with one unique frequency for each station. There is an interval of 0.2 s separating each of the eight transmissions, with variations in the duration for each station

*B. PFX Subsystem of the Akebono Satellite*

The PFX subsystem of the Akebono satellite measures 3 components of magnetic fields ($B_1$, $B_2$, and $B_3$) and 2 components of electric field ($E_x$ and $E_y$). It's comprised five-channel triple-super-heterodyne receivers with an output bandwidth of 50 Hz. It also had a local oscillator that could be stepped or fixed at a specific center frequency with a range from 100 Hz to 12.75 kHz and equipped with Wide Dynamic Range Amplifier (WIDA) hybrid IC to control the gain in the dynamic range for more than 80 dB [1]. WIDA hybrid IC will check automatically averaged signal level every 0.5 seconds and the gain of each channel is changed independently in 25 dB steps from 0 dB up to 75 dB. For our study, we use $B_X$, $B_Y$, $B_Z$, $E_X$ and $E_Y$ in static coordinate system converted from $B_1$, $B_2$, $B_3$, $E_x$ and $E_y$ obtained in the antenna coordinate system fixed to the spinning satellite as defined in Kimura et al. (1990). The static coordinate system is referred to the direction of the geomagnetic field (the geomagnetic field line is in the X-Z plane and Y is perpendicular to the X-Z plane) and the direction of the sun (Z-axis).

Fig. 2 shows the 10 s raw waveform of the PFX data at 18:05:29.503 UT on December 14, 1989. In the figure, we can observe the raised intensity from 18:05:33.500 UT to 18:05:34.900 UT, which in this case is the Omega signal from the Australia station. In the electric field components of $E_X$ and $E_Y$ from 18:05:33.800 UT to 18:05:34.300 UT, the PFX receivers both for $E_x$ and $E_y$ were saturated as indicated by the red arrows, because an intense Omega signal was captured. But the gain of these channels was immediately adopted adequately at 18:05:33.800 UT thanks to the WIDA IC. The WIDA IC worked independently for each five components. Therefore, all channels are not necessarily saturated but some components of the magnetic and/or electric field were occasionally saturated. During the period of saturation, signal intensity is apparently clipped and we cannot estimate absolute intensity. Then we define 'raise time' of Omega signal at the beginning point of the signal, while we use the data 0.5 seconds after the raise time when we evaluate 'absolute intensity' of the signals to exclude the saturated data. The detailed detection algorithm to derive raise time and intensity is described in the next section.

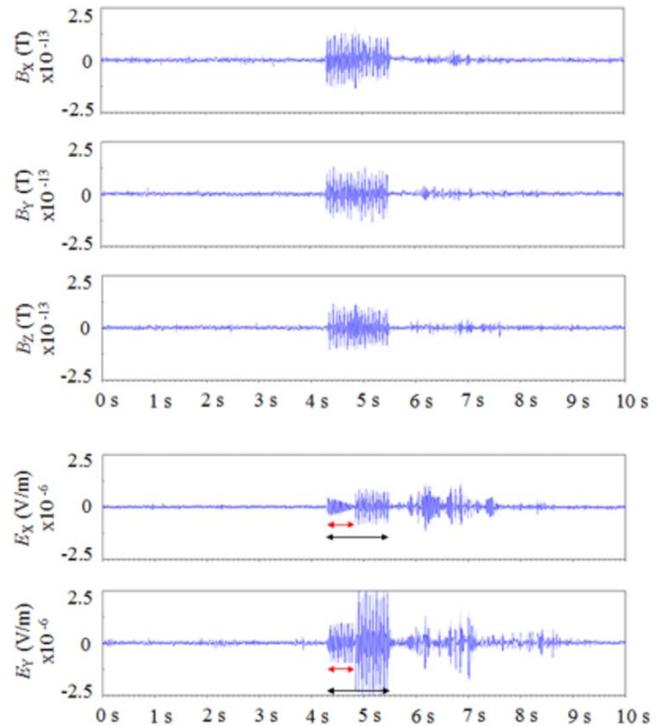

Fig. 2. Raw PFX waveform at 18:05:29.503 UT on December 14, 1989. We can observe the raised intensity from 18:05:33.500 UT to 18:05:34.900 UT, which in this case is the Omega signal from the Australia station. In the electric field components of $E_X$ and $E_Y$ from 18:05:33.800 UT to 18:05:34.300 UT, we can observe a 0.5 s saturated intensity signal caused by the WIDA IC

To measure Omega signals, we selected and analyzed the PFX data when its center frequency was fixed at 10.2 kHz. The PFX data was recorded in Common Data Format (CDF) developed by the National Space Science Data Center (NSSDC) at NASA [4]. This ensured standardized read/write interfaces for multiple programming languages and software. The waveforms measured by the PFX were originally two components of the electric field in the spin plane and three components of the magnetic field in the B1, B2 and B3 directions. These waveforms were orthogonal with respect to each other but different from satellite coordinates. These waveforms were calibrated, converted into static satellite coordinates, and stored in the CDF files. One month of PFX data represented 5–10 GB and one year of data consumed approximately 60–90 GB. In total, the amount of data from 1989 to 1997 is approximately 570 GB.

The PFX data is stored as waveforms sampled at rate of 320 Hz. For FFT analysis, we used FFT size of 32. Therefore, the time resolution was 100 ms and the frequency resolution was 10 Hz. To improve the accuracy of delay time detection, we applied an overlap-add FFT that moves over three sample points for a higher time resolution (~9.4 ms) when the first signal was detected. Although PFX measured only two components of the electric field, we could derive another component ($E_Z$) if we assumed the measured signal as a single plane wave [5]. This is expressed using (1).

$$E_Z = -\frac{E_X B_X + E_Y B_Y}{B_Z} \quad (1)$$

After we calculated the $E_Z$ component, we calculated the absolute intensity of the electric field $|E|$ as shown in (2). In the same way, we also calculated the absolute intensity of the magnetic field $|B|$ using (3). We use dB (V/m) for the electric field measurement unit and dB (T) for the magnetic field measurement unit.

$$|E| = \sqrt{E_X^2 + E_Y^2 + E_Z^2} \quad (2)$$

$$|B| = \sqrt{B_X^2 + B_Y^2 + B_Z^2} \quad (3)$$

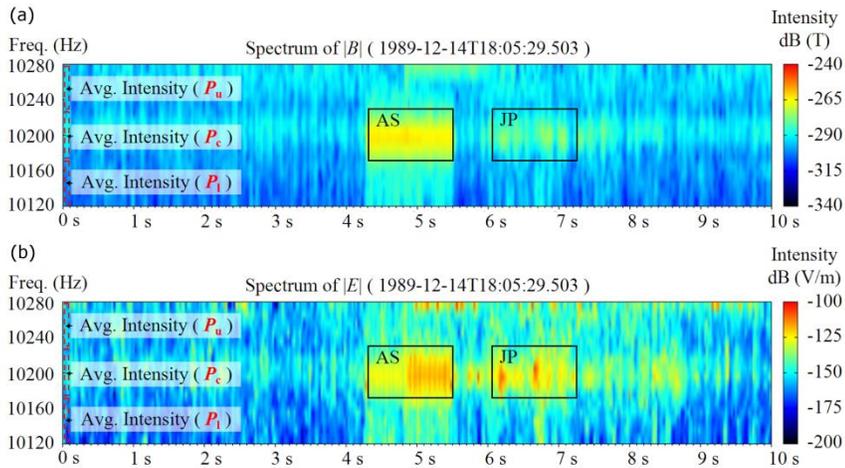

Fig. 3. Spectrum of the Omega signal start at 18:05:29.503 UT on December 14, 1989. We separated it into three frequency area bins ($P_c$, $P_u$, and $P_l$) to compare ambient noise and Omega signal intensity in the magnetic field (a) and in the electric field (b)

## III. DETECTION ALGORITHM

### A. Detection of the Omega Signal

We first estimated the raise time of each signal by comparing the average intensity of specific time frame to the threshold level, expecting a sudden increase in intensity. Second, we determined the transmission station by comparing the raise time with the transmission patterns of the eight Omega stations. At 0000 UT on January 1, 1972, the Omega and UTC scales were identical. However, we subsequently had to conduct a leap seconds calculation to synchronize the omega time and UTC because the Omega had no leap seconds like the UTC [5]. On December 31, 1989, the Omega time led the UTC by 14 s.

The threshold level we use in the detection method, is based on the comparison of ambient noise level and Omega signal intensity. In Fig. 3, we show parameters for the comparison of ambient noise and Omega signal intensity visualized on a spectrogram of 10 s of PFX data in the magnetic field (a) and in the electric field (b) beginning at 18:05:29.503 UT on December 14, 1989 when the Omega signal from Australia and Japan were expected to be received. The spectrogram in Fig. 3 consists of 16 bins in frequency (Δf = 10Hz/bin) and 1057 bins in time (Δt = ~9.4ms/bin). We separated it into three frequency area bins, where $P_c$ denotes the center frequency (consisting of 5 bins in frequency and 1 bin in time), $P_u$ denotes the upper frequency (consisting of 5 bins in frequency and 1 bin in time), and $P_l$ denotes the lower frequency (consisting of 6 bins in frequency and 1 bin in time).

Based on extracted data from the analyzer, Fig. 4 compares ambient noise and Omega signal intensity in the magnetic field at 18:05:29.503 UT on December 14, 1989. We can see the raised intensity level on $P_c$ compared with $P_u$ and $P_l$. This raised intensity level occurred within the expected transmission time of the Australia station (from approximately 18:05:33.700 UT to 18:05:34.900 UT) and the Japan station (from approximately 18:05:35.500 UT to 18:05:36.500 UT).

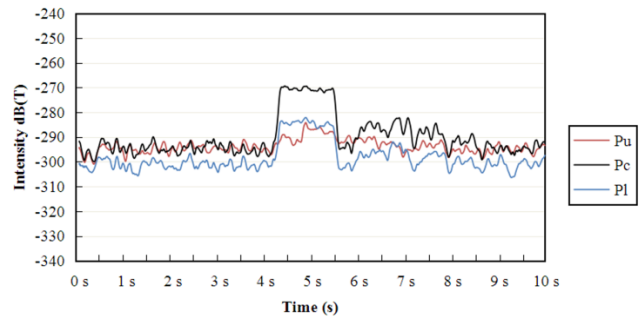

Fig. 4. Comparison of Omega signal intensity and ambient noise level in the magnetic field. Raised intensity level occurred within the expected transmission time of the Australia station (from approximately 18:05:33.700 UT to 18:05:34.900 UT) and the Japan station (from approximately 18:05:35.500 UT to 18:05:36.500 UT)

A comparison of ambient noise and omega signal intensity in the electric field can be seen in Fig. 5, which is based on extracted data from the analyzer at 18:05:29.503 UT on December 14, 1989. In this case, the signal was saturated from approximately 18:05:33.800 UT to 18:05:34.300 UT because the WIDA IC was controlling the gain of receiver. In this case, we need to calculate the intensity of the saturated signal after 0.5 s. We recognized this saturation by calculating and comparing each each signals for any sudden change in the intensity of the constant duration. In this case, the WIDA IC will affect the next 0.5 s sample for increased gain when activated. This type of saturated signal could affect any of the 5 components measured by the PFX subsystem because the WIDA IC works independently for each component.

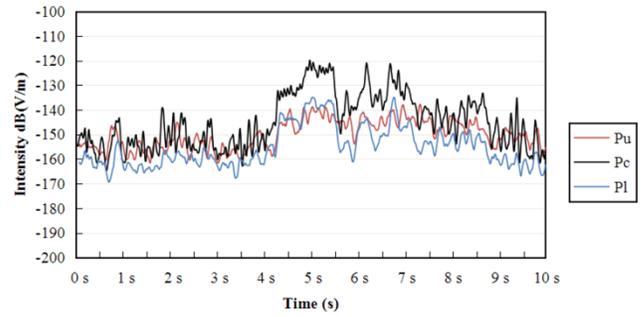

Fig. 5. Comparison of Omega signal intensity and ambient noise level in the electric field. The signal was saturated from approximately 18:05:33.800 UT to 18:05:34.300 UT because the WIDA IC was controlling the gain of receiver

### B. Calculation of Signal Intensity and Delay Time

We calculated the delay time of the signal by detecting the raise time of intensity and compare it with surrounding frequency from start time transmission during each station transmission duration as shown in (4).

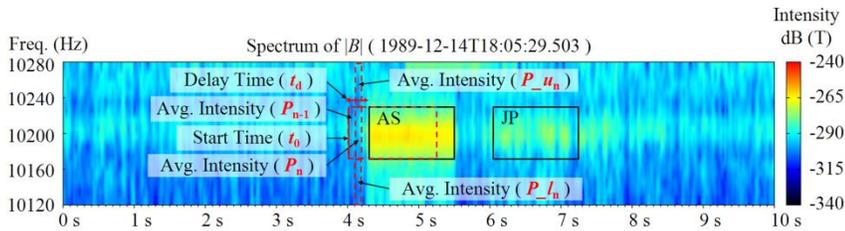

Fig. 6. Parameters used in the delay time detection method. Delay time ($t_d$) calculated by detecting the raise time of intensity ($P_n$) and compare it with surrounding frequency ($P\_u_n$ and $P\_l_n$) from start time transmission ($t_0$) during duration time of each station

$$t_d = \begin{cases} t_n - t_0 & \text{if } (P\_u_n + P\_l_n)/2 + T_P < P_n \text{ and } (P_n - P_{n-1}) > T_P \\ 0 & \text{otherwise} \end{cases} \quad (4)$$

where $T_P$ denotes the intensity threshold (8 dB), $t_d$ denotes delay time in seconds, $P_n$ denotes intensity strength of center frequency consists of 5 bins in frequency ($\Delta f$ = 10Hz/bin) and 20 bins in time ($\Delta t$ = ~9.4ms/bin), $P\_u_n$ denotes intensity strength of upper frequency consists of 5 bins in frequency ($\Delta f$ = 10Hz/bin) and 20 bins in time ($\Delta t$ = ~9.4ms/bin), $P\_l_n$ denotes intensity strength of lower frequency consists of 5 bins in frequency ($\Delta f$ = 10Hz/bin) and 20 bins in time ($\Delta t$ = ~9.4ms/bin), $t_0$ denotes start time of the station's transmission, $t_n$ denotes start time of $P_n$, and $n$ denotes the iteration number for every 1 bin in time ($\Delta t$ = ~9.4ms/bin). In Fig. 6, we show the parameters used for the delay time detection visualized on a spectrogram of 10 s of PFX data in the magnetic field start at 18:05:29.503 UT on December 14, 1989, when the Omega signal from Australia and Japan were expected to be received.

In the next step, we determined signal existence by comparing the intensity of the expected duration of the Omega signal with the surrounding intensity (higher and lower frequency points of the center frequency). We determined signal existence and derived the signal intensity ($P_{\omega s}$) by using (5)

$$P_{\omega s} = \begin{cases} P_s & \text{if } (P_a + P_b)/2 + T_P < P_s \text{ and } (P_w + T_w) < P_s \\ 0 & \text{otherwise} \end{cases} \quad (5)$$

where $P_a$ denotes the intensity strength of the upper frequency bins in decibels, $P_b$ denotes the intensity strength of lower frequency bins in decibels, $P_s$ denotes the intensity strength of center frequency bins in decibels. $P_a$, $P_b$, and $P_s$ consists of 5 bins in frequency ($\Delta f$ = 10Hz/bin) and variation (128 bins for Australia and 106 bins for Japan) of bins in time ($\Delta t$ = ~9.4ms/bin) depends on the duration of the Omega signal for each station. $P_w$ denotes intensity strength of 16 bins in frequency ($\Delta f$ = 10Hz/bin) and 1057 bins in time ($\Delta t$ = ~9.4ms/bin) for every 10 s duration (1 window) in decibels.

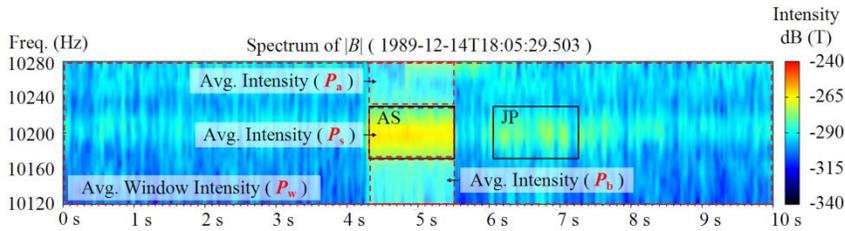

Fig. 7. Parameters used for discrimination and intensity calculations. The signal existence determined by comparing the average intensity of the expected duration of the Omega signal ($P_s$) with the surrounding frequency ($P_a$ and $P_b$) and the ambient noise of 10 seconds duration or 1 window ($P_w$)

In Fig. 7, we show the parameters used for the calculation of the signal visualized on the same spectrogram as shown in Fig. 6. In Fig. 7, for example, $P_s$ has total of 640 bins (5 bins in frequency and 128 bins in time) because the Omega signal from Australia station has 1.2 s in duration and we used overlap-add FFT. We defined two intensity thresholds $T_P$ and $T_W$, where $T_P$ had a fixed value of 8 dB and $T_W$ had a fixed value of 5 dB.

### C. PFX Analyzer

We developed software to analyze the Omega signal data measured by the Akebono from 1989 to 1997. The software was written in Java programming language. We can interactively check the waveform measured by the PFX one by one for an event study. It also enables us to detect Omega signals automatically for several months or years and then show the results of electric and magnetic intensity and delay time for specific locations of longitude, latitude, and altitude on geomagnetic and geographic maps of the Earth. Analyses of local time dependence are also available. An overview of the analyzer software is shown in Fig. 8.

This analyzer connects to the Akebono orbit database and can be used to manually analyze each signal in real time by manipulating the navigation panel. To automatically analyze the CDF files for one month of data, we need 6–9 hours of processing time using an Intel Quad Core with a 3.324 GHz CPU and 4 GB of memory. The computational time depends on the number of available signal data and data size. Depending on the computer specifications, it is possible to run multiple instances of the analyzer to speed up the process.

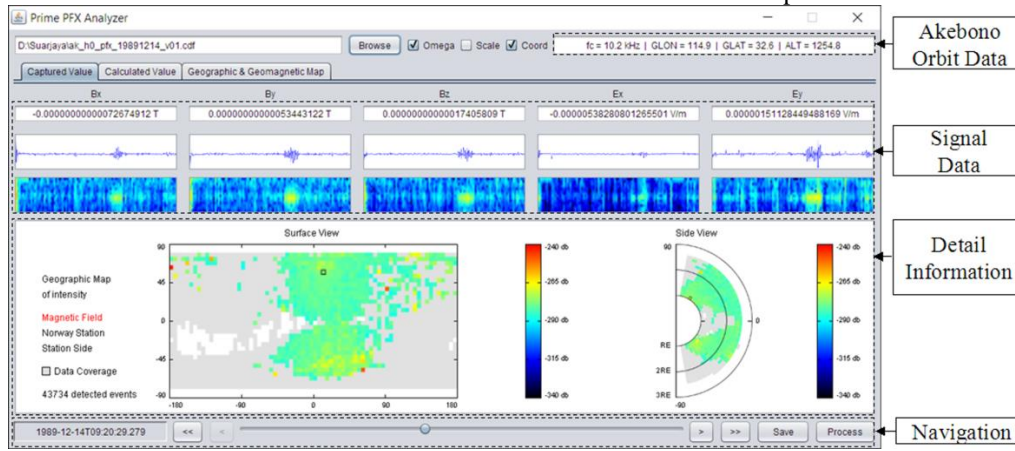

Fig. 8. Overview of the PFX Analyzer, written in Java programming language. It enables us to detect Omega signals automatically for several months or years and then show the results on geomagnetic and geographic maps of the Earth

### D. Handling of Error Detection and Evaluation of the System

For accurate results, we also applied error detection handling and a data smoothing process. It is possible for the analyzer to detect a high intensity noise as Omega signal. To handle this type of error detection, we applied a function that ignores a signal that is not detected continuously for a specific duration of time. This is currently set at 40 s. For example, when the signal is only detected for the first 30 s and then disappears in the next 10 s, the analyzer will ignore the signal and decide there is no signal detected for the entire 40 s. This handling occurs in the background of the analyzer software and the analyzer will still show a rectangle representing the detected signal.

We also applied a time smoothing algorithm to handle sudden peaks of delay time that may reflect false detection. This was accomplished by using the average of every two delay time value of detected signals. We expect to obtain more reliable results using improved detection data.

IV. RESULT OF ANALYSES

A. *Event Study*

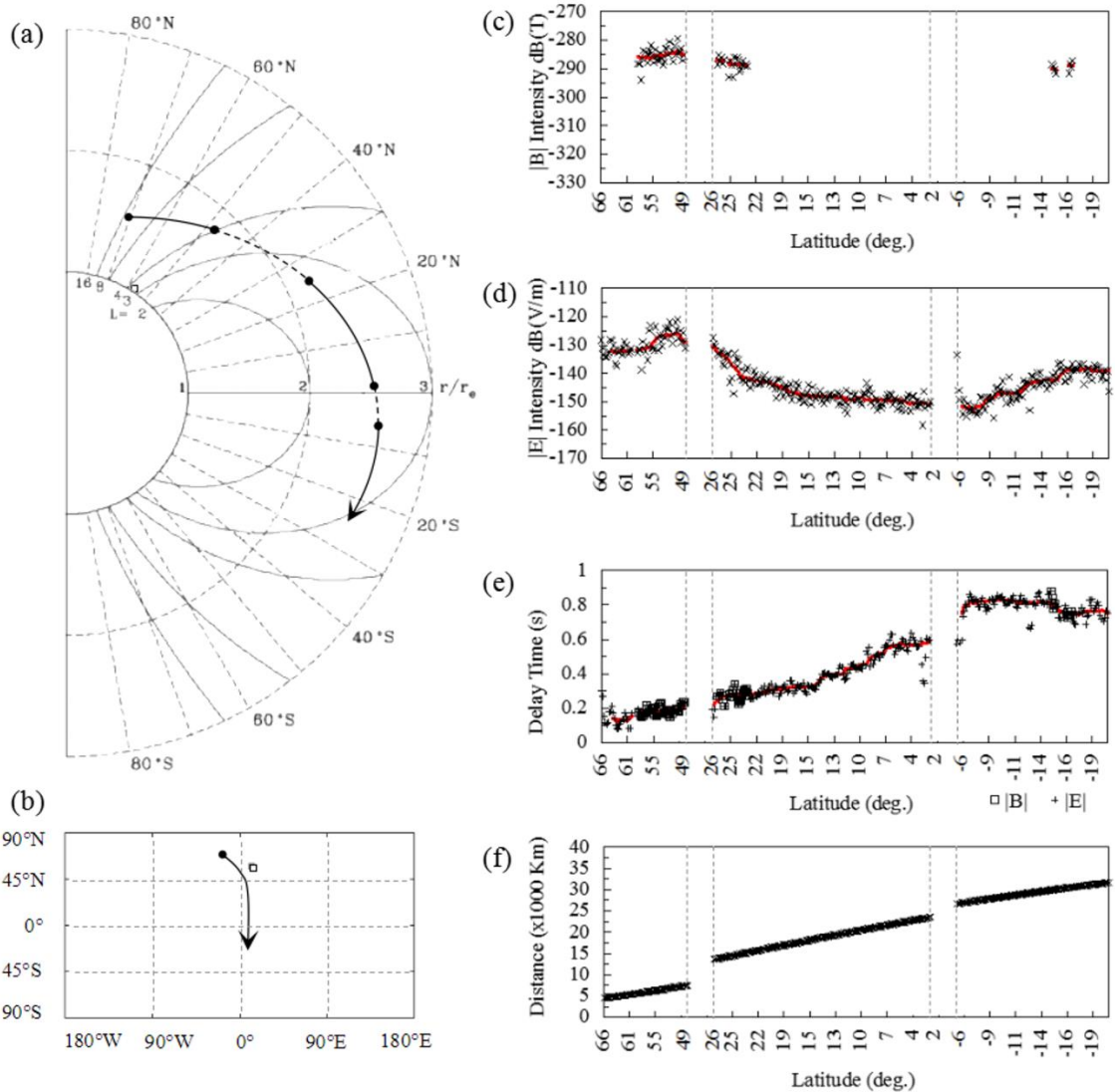

Fig. 9. Trajectory of Akebono, and observed Omega signal of Norway station from 08:29:39.802 UT to 09:55:33.842 UT on October 18, 1989. The PFX data is not available between latitude 49°N to 26°N and 2°N to 6°S and shown as dashed line. Left panels show trajectory line (a) (b) of the satellite. Right panels show (c) absolute intensities of Omega signal in the magnetic fields, (d) absolute intensities of Omega signal in the electric fields, (e) delay times of Omega signal, and (f) the approximate distance of the propagation path to the observation point

Fig. 9 (a) and 9 (b) shows 1.5 hours of trajectory of the Akebono from the northern to southern hemisphere from 08:29:39.802 UT to 09:55:33.842 UT on October 18, 1989. During this period, Omega signals from the Norway station were continuously observed and 344 signals were detected in the electric field and 63 signals were detected in the magnetic field. The location of the Norway station is also shown as a small square in Fig. 9 (a) and (b). The PFX data is not available between latitude 49°N to 26°N and 2°N to 6°S. This unavailable of data is shown as dashed lines in Fig. 9 (a).

Fig. 9 (c) shows absolute intensities of Omega signal in the magnetic fields, Fig. 9 (d) shows absolute intensities of Omega signal in the electric fields, and Fig. 9 (e) shows delay time of the Omega signals. The red curves indicate moving median over 20 points (detected events). The intensity of the magnetic field and electric field of the Omega signal is showing higher intensity in the northern hemisphere where the Norway station was located. The Norway station was located at latitude 56.42°N, and we can see higher signal intensity around −285 dB (T) for the magnetic field and approximately −125 dB (V/m) for the electrical field. The intensity then becomes lower for the electric field or disappears for the magnetic field near the equator at latitude 0°. The signal for the electric field then increased at approximately −145 dB (V/m) near the southern hemisphere because the signal could propagate along the Earth's magnetic field. The delay time around 0.2 s in the northern hemisphere where the Norway station was located and then increased to more than 0.5 s as the trajectory of the Akebono satellite got closer to the southern hemisphere. This occurred because propagation along the Earth's magnetic field required more time compared with a direct propagation path. Fig. 9 (f) shows the approximate distance based on a simple magnetic dipole model for the propagation path of the signal along the Earth's magnetic field. The calculated distance was from the Norway station to the observation point.

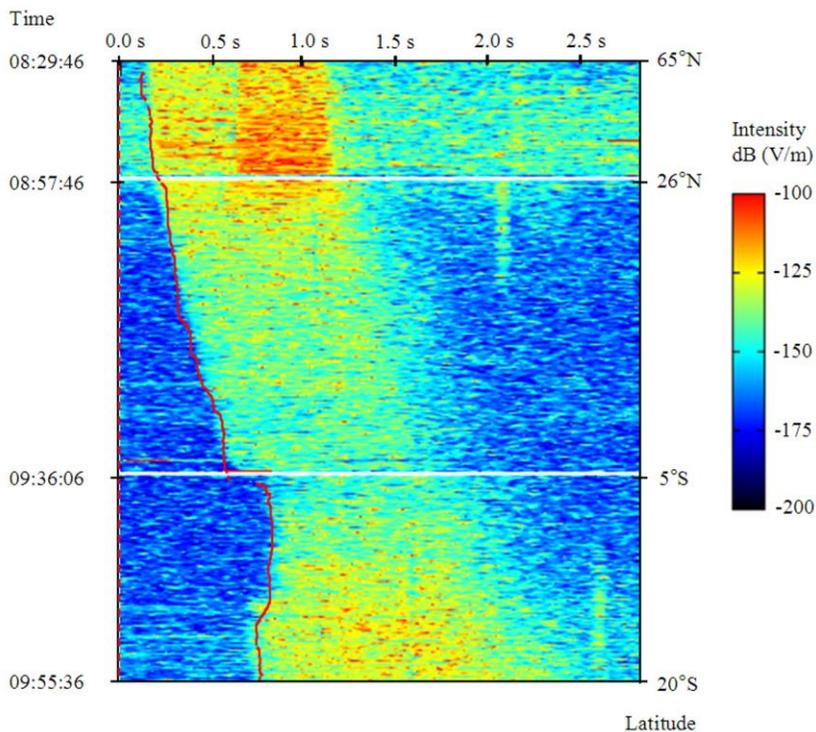

Fig. 10. Sliced Center Frequency in the electric field from 08:29:46 UT to 09:55:36 UT on October 18, 1989. The transmission time of Norway station was 08:29:46 UT and repeated every 10 seconds. Signal from Norway expected to be 0.9 second in duration. Intensification at t=0.6 in a period from 08:29:46 UT to 08:57:46 UT was caused by the WIDA IC

For comparison, we calculated the sliced center frequency of each signal in the Electric Field from 08:29:46 UT to 09:55:36 UT on October 18, 1989. We plotted only 1 bin ($\Delta f = 10$ Hz/bin, $\Delta t = \sim 9.4$ ms/bin) of the center frequency spectrum ($\sim 10.2$ kHz) and arranged it as shown in the Fig. 10. The transmission time of Norway station start at 08:29:46 UT and repeated every 10 seconds. We projected the red curves of trend line in Fig. 9 (e) to Fig. 10. This confirms that the automatic detection of our analyzer produced quite reliable results compared with manual analysis. Deviation occurred at few parts because of false result by multiple high intensity noises and affected moving median of the trend line. Based on the Omega system transmission pattern, we expected signal from Norway should be 0.9 second in duration. Intensification at t=0.6 in a period from 08:29:46 UT to 08:57:46 UT was caused by the WIDA IC. Because of its weakness, WIDA IC may cause false intensification of signal as seen at t=2.05 in a period from 08:57:46 UT to 09:36:06 UT. The effects of WIDA IC can be seen clearly in Fig. 6. Signal from Liberia station was expected to be delayed and might be received after t=1.0. In period from 09:36:06 UT to 09:55:36 UT, overlapped signal between Norway and Liberia is possible.

*B. Statistical Study*

The results of more than three years data of Norway station (October 1989 to December 1992) is shown in Fig. 11. The longitudinal axis consists of 72 bins of 5° for each bin; the latitudinal axis consist of 36 bins of 5° for each bin; and the altitudinal axis consist of 20 bins of 637.1 Km for each bin. The rectangle on the map shows the location of the transmission station at a longitude of 13.14°E and latitude of 56.42°N. The gray color bin on the map shows the availability of PFX data in the area. However, no Omega signal was detected by the analyzer. The white color bin on the map shows that the PFX data is not available for that location. The meridian plane map shows the average intensity coverage of 10° to the east and 10° to the west from the station. From the figure, We can clearly demonstrate that Omega signal propagated nearly along the geomagnetic field line to the southern hemisphere. The intensity of the signal near the equator shows low intensity at high altitude and no signal at lower altitude. We concluded that this unique propagation is caused primarily by the location of the transmission station, the Earth's magnetic field, and global electron density. Previous study about global electron density had analyzed normal wave direction and delay time of the Omega signal [7][8] and also deduced from whistlers [9][10][11]. For more than three years (39 months), we observed 43,734 detected signals in the magnetic field and 111,049 detected signals in the electric field.

V. CONCLUSIONS

In this study, we developed an advanced detection algorithm to continuously process large amounts of data measured for several years by the PFX subsystem on board the Akebono satellite. The algorithm enables us to distinguish noise and real omega signals and also detect errors in order to produce more accurate results. When compared with manual analysis, automatic detection can be accomplished with short periods of processing time and does not require human intervention in the process. We demonstrated that the proposed method is powerful enough for the statistical analyses and further study of the propagation patterns of VLF waves with regard to the Earth's magnetic field and global electron density.

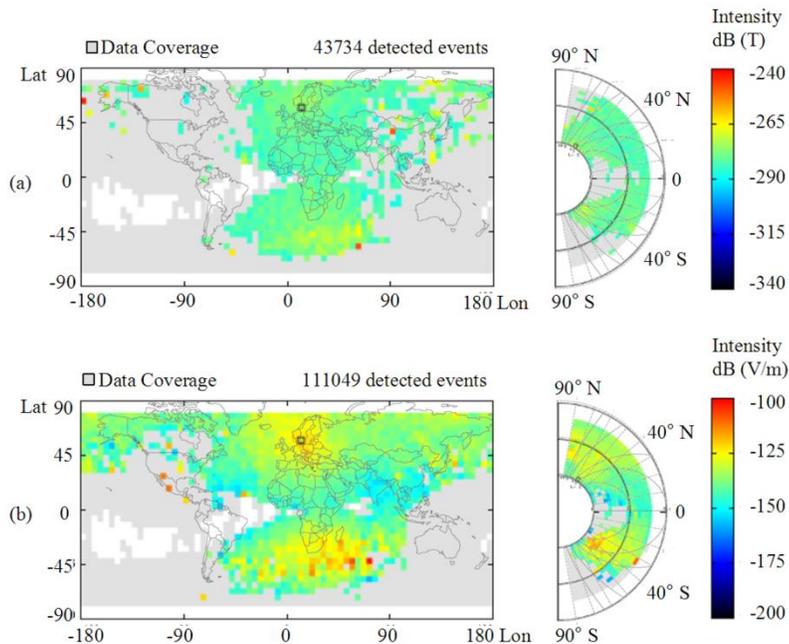

Fig. 11. Omega Signal Propagation of Norway Station from October 1989 to December 1992 in the magnetic field (a) and in the electric field (b). Omega signal is propagated to the southern hemisphere. The intensity of the signal near the equator shows low intensity at high altitude and no signal at lower altitude

ACKNOWLEDGMENT

This research was partially supported by a Grant-in-Aid for Scientific Research from the Japan Society for the Promotion of Science (#24360159 and #16H01172).